\documentclass{siamltex}
\usepackage{ae}
\usepackage{aecompl}
\usepackage[T1]{fontenc}
\usepackage[latin1]{inputenc}
\usepackage{float}
\usepackage{amsmath}
\usepackage{graphicx}
\usepackage{amssymb}

\makeatletter


\makeatother
\begin{document}

\title{Localized dispersive states in nonlinear coupled mode equations for
light propagation in fiber Bragg gratings%
\thanks{This work was supported by Spanish Direcci�n General de Investigaci�n
under grant MTM2004-03808 and by the Universidad Polit�cnica de Madrid
under grant CCG06-UPM/IME-328.%
}}

\author{C. Martel%
\thanks{Depto. de Fundamentos Matem�ticos, E.T.S.I. Aeron�uticos, Universidad
Polit�cnica de Madrid, Plaza Cardenal Cisneros 3, 28040 Madrid, SPAIN
(\texttt{martel@fmetsia.upm.es, maria@fmetsia.upm.es, carrasco@fmetsia.upm.es}).%
}\and M. Higuera$^{\dagger}$\and J.D. Carrasco$^{\dagger}$}

\maketitle
\begin{abstract}
Dispersion effects induce new instabilities and dynamics in the weakly
nonlinear description of light propagation in fiber Bragg gratings.
A new family of dispersive localized pulses that propagate with the
group velocity is numerically found and its stability is also analyzed.
The unavoidable different asymptotic order of transport and dispersion
effects plays a crucial role in the determination of these localized
states. These results are also interesting from the point of view
of general pattern formation since this asymptotic imbalance is a
generic situation in any transport dominated (i.e., nonzero group
velocity) spatially extended system.
\end{abstract}

\section{Introduction}

Fiber Bragg gratings (FBG) are microstructured optical fibers that
present a spatially periodic variation of the refractive index. The
combination of the guiding properties of the periodic media with the
Kerr nonlinearity of the fiber results in the very particular light
propagation characteristic of these elements, which make them very
promising for many technological applications that range from optical
communications (wavelength division, dispersion management, optical
buffers and storing devices, etc.) to fiber sensing (structural stress
measure in aircraft components and buildings, temperature change detection,
etc.), see, e.g., the recent review \cite{Kashyap99}.

The amplitude equations that are commonly used in the literature to
model one dimensional light propagation in a FBG are the so-called
nonlinear coupled mode equations (NLCME) \cite{WinfulCooperman82,deSterkeSipe94,deSterke98,Aceves00,GoodmanWeinsteinHolmes01},
\textbf{}which, conveniently scaled,can be written as\begin{eqnarray}
A_{t}^{+}-A_{x}^{+}=\textrm{i}\kappa A^{-}+\textrm{i}A^{+}(\sigma|A^{+}|^{2}+|A^{-}|^{2}),\label{eq:NLCME+}\\
A_{t}^{-}+A_{x}^{-}=\textrm{i}\kappa A^{+}+\textrm{i}A^{-}(\sigma|A^{-}|^{2}+|A^{+}|^{2}),\label{eq:NLCME-}\end{eqnarray}
 where $A^{\pm}$ are the envelopes of the two counterpropagating
wavetrains that resonate with the grating, $\kappa$ is the strength
of the coupling effect produced by the grating and $\sigma>0$ is
ratio of the self to cross nonlinear interaction coefficient ($\sigma=\frac{1}{2}$
for a cubic Kerr nonlinearity \cite{deSterkeSipe94}). The NLCME can
be obtained from the full Maxwell-Lorentz equations using multiple
scales techniques in the limit of small grating depth, small light
intensity and slow spatial and temporal dependence of the field envelopes
(see \cite{GoodmanWeinsteinHolmes01} for a detailed description of
this derivation process).

It has been recently shown \cite{Martel05,MartelCasas06} that light
propagation in FBG can develop dispersive structures that are not
accounted for in the NLCME formulation, and that, to correctly describe
the weakly nonlinear dynamics of the system, the NLCME have to be
completed with material dispersion terms:\begin{eqnarray}
A_{t}^{+}-A_{x}^{+}=\textrm{i}\kappa A^{-}+\textrm{i}A^{+}(\sigma|A^{+}|^{2}+|A^{-}|^{2})+\textrm{i}\varepsilon A_{xx}^{+},\label{eq:NLCMEd+}\\
A_{t}^{-}+A_{x}^{-}=\textrm{i}\kappa A^{+}+\textrm{i}A^{-}(\sigma|A^{-}|^{2}+|A^{+}|^{2})+\textrm{i}\varepsilon A_{xx}^{-}.\label{eq:NLCMEd-}\end{eqnarray}
 The dispersive nonlinear coupled mode equations above (NLCMEd) are
scaled as the NLCME: the characteristic length scale is the slow scale
that results from the balance of the advection term with the small
effect of the grating, the characteristic time is the corresponding
transport time scale (which sets to one the scaled group velocity),
and the characteristic size of the wavetrains is the resulting one
from the saturation of the small nonlinear terms. The slow envelope
assumption forces the dispersive terms to be always small as compared
with the advection terms; in other words, second derivatives of the
slow amplitudes are much smaller than first derivatives. In the scaled
equations this effect is contained in the scaled dispersion coefficient
$\varepsilon$ (which measures the dispersion to transport ratio)
and therefore, in order to be consistent with the slow envelope assumption,
the NLCMEd must be considered in the limit $\varepsilon\rightarrow0$.

The NLCMEd can be somehow regarded as asymptotically nonuniform, in
the sense that the NLCMEd is an asymptotic model obtained in the $\varepsilon\rightarrow0$
(weakly nonlinear, slow envelope) limit that still contains the small
parameter $\varepsilon$. This is the unavoidable consequence of simultaneously
considering two balances of different asymptotic order: one induced
by the dominant effect of the transport at the group velocity (balance
described by NLCME) and a second one that is associated with the underlying
dispersive, nonlinear Schr�dinger-like dynamics of the system. The
small dispersive terms in the NLCMEd are essential to describe the
dynamics of the system when it develops small dispersive scales $\delta_{\textrm{disp}}\sim\sqrt{|\varepsilon|}$.
As it was shown in \cite{Martel05}, the NLCMEd in the $\varepsilon\rightarrow0$
limit constitute a singular perturbation problem (cf. \cite{KevorkianCole96})
and the onset of the dispersive scales is not a higher order, longer
time effect; it takes place in the same timescale of the NLCME, no
matter how small is the dispersion coefficient $\epsilon$.

\begin{figure}
\begin{centering}
\includegraphics[bb=0bp 0bp 503.00bp 324bp,scale=0.73]{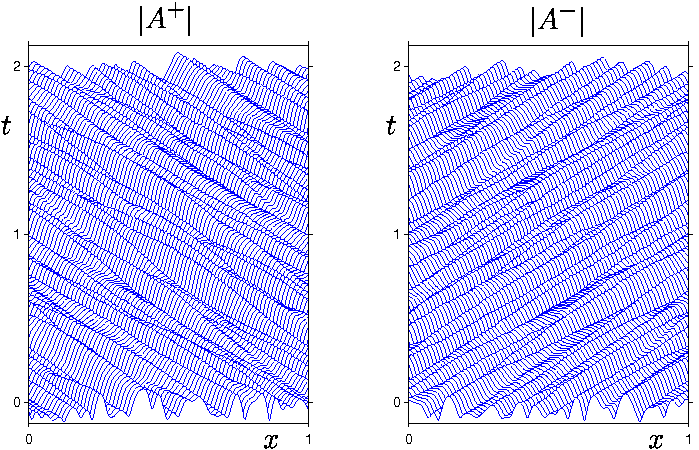}
\par\end{centering}

\caption{\label{fig:xtchaos1}Space-time representation of a solution of the
NLCMEd exhibiting small dispersive scales all over the domain ($\sigma=1/2$,
$\kappa=2$, $\varepsilon=-10^{-3}$, and periodicity boundary conditions).
See the file \texttt{movie1.1.gif} for an animation of the onset of
the dispersive scales.}
\end{figure}

\begin{figure}
\begin{centering}
\includegraphics[bb=0bp 0bp 503bp 324bp,scale=0.73]{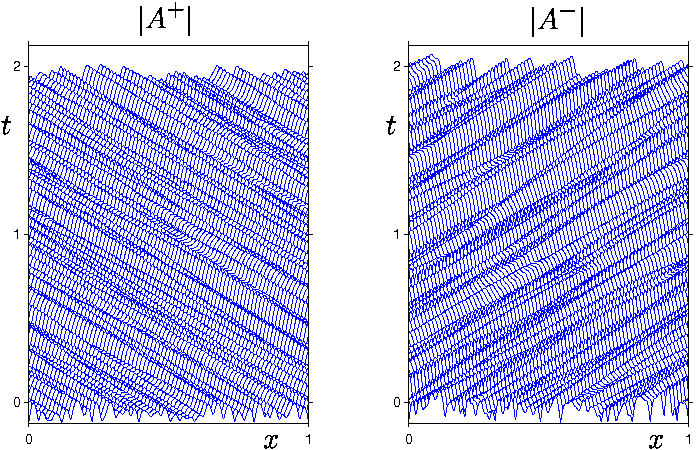}
\par\end{centering}

\caption{\label{fig:xtchaos2}Space-time representation of a solution of the
NLCMEd exhibiting small dispersive scales all over the domain ($\sigma=1/2$,
$\kappa=2$, $\varepsilon=-10^{-3}/4$, and periodicity boundary conditions).
See the file \texttt{movie1.2.gif} for an animation of the onset of
the dispersive scales.}
\end{figure}

A solution of the NLCMEd for $\varepsilon=-10^{-3}$ that exhibits
small dispersive scales all over the domain is represented in Figure~\ref{fig:xtchaos1}:
note that, for short time, the dispersive structures just propagate
with the group velocity but, for $t\sim1$, they also interact with
each other giving rise to a very complicated spatio-temporal pattern.
The initial condition used in this simulation was a uniform modulus
solution with a small random perturbation that, according to the dispersion-less
NLCME formulation, was a stable solution. In order ensure that the
small scales in Figure~\ref{fig:xtchaos1} are dispersion induced
scales we have repeated the NLCMEd simulation but with a reduced dispersion
$\varepsilon=-10^{-3}/4$. The result is plotted in Figure~\ref{fig:xtchaos2}
where it can be seen that the small scales fill again the entire domain,
but its typical size, $\delta_{\textrm{disp}}\sim\sqrt{|\varepsilon|}$,
is now approximately one half of that in Figure~\ref{fig:xtchaos1}
(see also the corresponding animations \texttt{movie1.1.gif} and \texttt{movie1.2.gif}).

The main goal of this paper is to show that,  in addition to complex
spatio-temporal patterns, the dispersion effects can also give rise
to new, purely dispersive localized states, which might be of interest
from the optical communications point of view. It is interesting to
note that the results in this paper apply also to Bose-Einstein condensates
in optical lattices (a system that has recently received very much
attention \cite{SakaguchiMalomed04,yulinskryabin03}) and, in general,
to any dissipation-less propagative system, extended in one spatial
direction, reflection and translation invariant, and with a small
superimposed spatial periodic modulation of its background, since
the NLCMEd are the appropriate envelope equations for the description
of the weakly nonlinear resonant dynamics of this kind of systems. 

In order to show that light propagation in a FBG can happen in the
form of dispersive pulses, we derive and solve numerically in section
2 an asymptotic equation for a family of symmetric pulses, and, in
section 3, we perform some numerical integrations of the complete
NLCMEd to show that some of the pulses in this family do propagate
as stable localized structures. Finally, some concluding remarks are
drawn in section 4.

\section{Dispersive pulses}

The starting point is the continuous wave (CW) family of constant
uniform modulus solutions of the NLCME (\ref{eq:NLCME+})-(\ref{eq:NLCME-})

\begin{eqnarray}
 & A_{\textrm{CW}}^{+}=\rho\cos\theta\: e^{i\omega t+imx},\label{CW+}\\
 & A_{\textrm{CW}}^{-}=\rho\sin\theta\: e^{i\omega t+imx},\label{CW-}\end{eqnarray}
 where $\rho\ge0$ is the light intensity flowing through the fiber,
$\theta\in[-\frac{\pi}{2},\frac{\pi}{2}]$ measures the relative amount
of both wavetrains, and the frequency and wavenumber of the amplitudes
are given by\begin{eqnarray}
 & \omega=\dfrac{\kappa}{\sin2\theta}+\dfrac{\sigma+1}{2}\rho^{2},\label{alfa}\\
 & m=(\dfrac{\kappa}{\sin2\theta}-\dfrac{\sigma-1}{2}\rho^{2})\cos2\theta.\label{eme}\end{eqnarray}
 The CW with $|\omega|\sim1$ and $|m|\sim1$ are approximate solutions
of the NLCMEd (up to order $\varepsilon$ corrections) and its stability
was first analyzed in \cite{deSterke98} and then completed in \cite{Martel05},
where it was found that, for both signs of the dispersion coefficient,
there are dispersively unstable CW which are stable in the dispersion
less context of the NLCME.

\begin{figure}
\begin{centering}
\includegraphics[bb=0bp 0bp 503bp 324bp,scale=0.4]{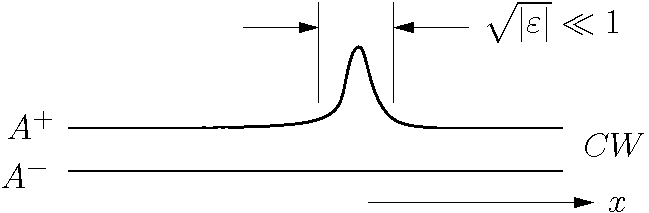}
\par\end{centering}

\caption{\label{fig:sketchpulse}Sketch of a dispersive pulse on top of a
continuous wave.}
\end{figure}

We now look for localized dispersive pulses propagating on top of
one of the amplitudes of a stable CW, as sketched in Figure~\ref{fig:sketchpulse}.
In order to turn the background CW into a constant, it is convenient
to first perform in the NLCMEd the following change of variables\begin{eqnarray*}
 & A^{+}=F^{+}\: e^{i\omega t+imx},\\
 & A^{-}=F^{-}\: e^{i\omega t+imx},\end{eqnarray*}
 to obtain\begin{eqnarray}
F_{t}^{+}-F_{x}^{+}+\textrm{i}(\omega-m)F^{+}=\textrm{i}\kappa F^{-}+\textrm{i}F^{+}(\sigma|F^{+}|^{2}+|F^{-}|^{2})+\textrm{i}\varepsilon F_{xx}^{+},\label{eq:F+}\\
F_{t}^{-}+F_{x}^{-}+\textrm{i}(\omega+m)F^{-}=\textrm{i}\kappa F^{+}+\textrm{i}F^{-}(\sigma|F^{-}|^{2}+|F^{+}|^{2})+\textrm{i}\varepsilon F_{xx}^{-}.\label{eq:F-}\end{eqnarray}
 A localized dispersive pulse on $F^{+}$ depends on the fast spatial
scale $X=x/\sqrt{|\varepsilon|}$ and, according to (\ref{eq:F+})-(\ref{eq:F-}),
in the short time scale, $T=t/\sqrt{|\varepsilon|}\sim1$, it just
propagates with the group velocity, suggesting that we have to look
for solutions of the form \begin{eqnarray*}
 & F^{+}=F_{0}^{+}(\eta,t)+\cdots,\\
 & F^{-}=F_{0}^{-}(\eta,t)+\dots,\end{eqnarray*}
 with $\eta=X+T$. Inserting the above ansatz into (\ref{eq:F+})-(\ref{eq:F-})
yields\[
F_{0\eta}^{-}=0,\quad\textrm{that gives}\quad F_{0}^{-}=\rho\sin\theta,\]
 which means that $F^{-}$ remains in first approximation equal to
the unperturbed CW. Similarly, the following equation is obtained
for $F_{0}^{+}$\begin{equation}
F_{0t}^{+}+\textrm{i}(\omega-m)F_{0}^{+}=\textrm{i}\kappa\rho\sin\theta+\textrm{i}F_{0}^{+}(\sigma|F_{0}^{+}|^{2}+\rho^{2}\sin^{2}\theta)\pm\textrm{i}F_{0\eta\eta}^{+},\label{eq:F0+}\end{equation}
 where the $+$ ($-$) sign corresponds to $\varepsilon$ positive
(negative), together with the boundary conditions\begin{equation}
F_{0}^{+}\rightarrow\rho\cos\theta\quad\textrm{for}\quad\eta\rightarrow\pm\infty,\label{eq:F0+bc}\end{equation}
 which ensure that the background CW is recovered away from the pulse.

For dispersive pulses propagating over a zero background we have to
set to zero $\rho$, $\omega$ and $m$ in eq. (\ref{eq:F0+}). A
standard nonlinear Schr�dinger (NLS) equation is then obtained, which
is known to exhibit localized pulses (solitons) in the focusing case
of positive dispersion (recall that $\sigma>0$). Note that in this
case the effect of the grating is completely gone because it takes
place only through the background state. The only nonzero Fourier
spectrum components of these NLS solitons correspond, in first approximation,
to very large dispersive wavenumbers ($\sim1/\sqrt{|\varepsilon|}\gg1$)
that are so off-resonance that they simply do not feel the grating.

Equation (\ref{eq:F0+}) is a nonlinear Schr�dinger equation with
a direct forcing term coming from the effect of the grating. The steady
solutions of this equation and their stability properties were analyzed
in \cite{Barashenkov96,Barashenkov89}, where explicit analytic expressions
for the steady pulses were found. In this paper we consider the more
general family of traveling localized solutions. More precisely, we
look for traveling pulses of the form\[
F_{0}^{+}=\rho\cos\theta(1+a(\eta+vt)),\]
 where $v$ represents, in the original variables, a small correction
of the group velocity. The resulting boundary value problem for $a$,
after making use of relations (\ref{alfa}) and (\ref{eme}) and the
rescaled variable $\xi=(\eta+vt)(\sqrt{\sigma}\rho\cos\theta)$, can
be written as \begin{eqnarray}
a_{\xi\xi} & =-i\hat{v}a_{\xi}+\alpha a-(|a|^{2}+a+\bar{a})(1+a),\label{eq:eqp2}\\
 & a\rightarrow0\text{\quad as\quad}\xi\rightarrow\pm\infty\label{eq:eqbcp2}\end{eqnarray}
 where $\hat{v}=v/(\sqrt{\sigma}\rho\cos\theta)$ and $\alpha=k\tan(\theta)/(\sigma\rho^{2}\cos^{2}\theta)$.
We are restricting our search to the focusing case of positive sign
in eq. (\ref{eq:F0+}) and, in order to have a dispersively stable
background CW, we have to consider only the range $0<\theta<\frac{\pi}{2}$
(see \cite{Martel05}), which implies that we have to look for pulses
in eqs. (\ref{eq:eqp2})-(\ref{eq:eqbcp2}) only for $\alpha>0$.
On the other hand, eqs. (\ref{eq:eqp2})-(\ref{eq:eqbcp2}) remain
invariant under the transformation $\hat{v}\rightarrow-\hat{v}$ and
$a\rightarrow\bar{a}$, and therefore we can also set $\hat{v}\ge0$.

\begin{figure}
\begin{centering}
\includegraphics[bb=0bp 0bp 503bp 324bp,scale=0.4]{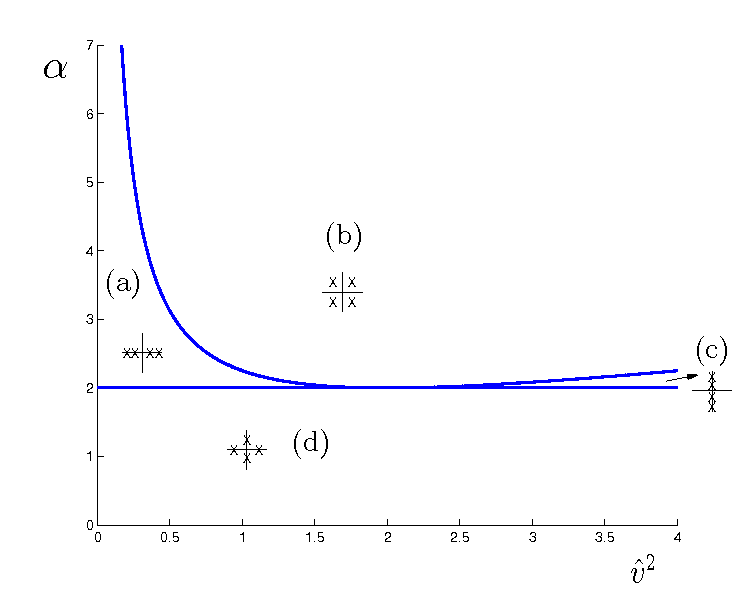}
\par\end{centering}

\caption{\label{fig:eigval}Eigenvalues of the trivial state as a function
of the parameters $\alpha$ and $\hat{v}^{2}$. Sketches show the
distribution of the four eigenvalues in each of the distinct regions.}
\end{figure}

The solutions of eqs.~(\ref{eq:eqp2})-(\ref{eq:eqbcp2}) correspond
to traveling dispersive pulses and can be regarded as homoclinic orbits
(in the variable $\xi$) connecting the trivial state (i.e., $a=0$)
back to itself. 

To analyze the existence of such connections we first consider the
linearized system around the zero state:\textbf{\begin{equation}
\frac{d\bold u}{d\xi}=A_{\infty}\bold u\label{eq:lin}\end{equation}
}where ${\bold u}^{T}\equiv(a_{r},a_{i},(a_{r})_{\xi},(a_{i})_{\xi})$,
$a_{r}$ and $a_{i}$ are the real and imaginary part of $a$, $(a_{r})_{\xi}=da_{r}/d\xi$
and $(a_{i})_{\xi}=da_{i}/d\xi$, and the matrix $A_{\infty}$ is
given by\textbf{\begin{equation}
A_{\infty}=\left(\begin{array}{cccc}
0 & 0 & 1 & 0\\
0 & 0 & 0 & 1\\
\alpha-2 & 0 & 0 & \hat{v}\\
0 & \alpha & -\hat{v} & 0\end{array}\right).\label{eq:Ainfty}\end{equation}
}The eigenvalues of this system are given by $\lambda_{\pm}^{1}=\sqrt{\eta_{\pm}}$
and $\lambda_{\pm}^{2}=-\sqrt{\eta_{\pm}}$, with \begin{equation}
\eta_{\pm}=\left[2\alpha-(\hat{v}^{2}+2)\pm\sqrt{(\hat{v}^{2}+2)^{2}-4\hat{v}^{2}\alpha}\right]/2.\label{eq:eqeig}\end{equation}
Fig.~\ref{fig:eigval} shows the behaviour of the four eigenvalues
$\lambda_{\pm}^{1}$ and $\lambda_{\pm}^{2}$ in the $(\alpha,\hat{v}^{2})$
plane. There are four distinct regions, separated by the boundaries
$\alpha=2$ and $\alpha=(\hat{v}^{2}+2)^{2}/4\hat{v}^{2}$: (a) four
real eigenvalues, (b) two pairs of complex conjugate eigenvalues,
(c) four purely imaginary eigenvalues and (d) two real eigenvalues
along with two purely imaginary eigenvalues. In regions (a) and (b)
the unstable and stable manifolds of the origin are two dimensional,
while the equilibrium is nonhyperbolic in regions (c) and (d) where
the center manifold is, respectively, four- and two-dimensional. Homoclinic
orbits belong to both the stable and unstable manifold of the origin.
We investigate below the presence of homoclinic solutions in cases
(a) and (b), where these correspond to the intersections of two dimensional
stable and unstable manifolds in a four dimensional space \cite{Wiggins88}.

The problem (\ref{eq:eqp2})-(\ref{eq:eqbcp2}) is invariant under
the symmetry \begin{equation}
a\rightarrow\bar{a},\qquad\xi\rightarrow-\xi,\label{eq:eqsym}\end{equation}
that comes from the time reversing (Hamiltonian) and spatial reflection
symmetries of the NLCMEd. We further restrict our search for dispersive
pulses to the case of reflection-symmetric pulses, i.e., to pulses
that satisfy:\[
a(\xi)=\bar{a}(-\xi).\]
If we now set the symmetry axis to $\xi=0$, we can reduce the problem
to a semi-infinite interval $\xi\in[0,+\infty[$ with the boundary
conditions \begin{eqnarray}
(a_{r},a_{i}) & = & (a_{0},0)\quad\text{and}\quad((a_{r})_{\xi},(a_{i})_{\xi})=(0,b_{0}),\quad\text{at}\quad\xi=0,\label{eq:eqbc0}\\
(a_{r},a_{i}) & \rightarrow & (0,0)\quad\text{as}\quad\xi\rightarrow\infty.\label{eq:eqbcinf}\end{eqnarray}

Finally, to numerically compute the profiles of the symmetric pulses,
we replace the infinite interval by a finite one, $[0,L]$. Following
\cite{Keller76}, the resulting boundary conditions at $\xi=L$ are
obtained by requiring that the solution projects only onto the subspace
spanned by the eigenvectors associated with the decaying eigenvalues
of the matrix $A_{\infty}$ (see eq. (\ref{eq:lin})). In summary,
the boundary value problem that we integrate numerically is given
by eq.~(\ref{eq:eqp2}), which we rewrite as a real first order system
of four equations in $]0,L[$ together with the four boundary conditions
\begin{eqnarray}
C_{0}\bold u & = & \bold0\quad\text{at}\quad\xi=0,\quad\text{and}\quad\label{eq:bndrycond0}\\
C_{\infty}\bold{u} & = & \bold0\quad\text{at}\quad\xi=L,\label{eq:bndrycondinf}\end{eqnarray}
where ${\bold u}^{T}\equiv(a_{r},a_{i},(a_{r})_{\xi},(a_{i})_{\xi})$,
\textbf{\begin{equation}
C_{0}=\left(\begin{array}{cccc}
0 & 1 & 0 & 0\\
0 & 0 & 1 & 0\end{array}\right),\label{eq:C0}\end{equation}
}and $C_{\infty}$ is a matrix whose rows are the left eigenvectors
of $A_{\infty}$ associated with the exponentially growing directions\textbf{\begin{equation}
C_{\infty}=\left(\begin{array}{cccc}
(\eta_{+}-\alpha-\hat{v}^{2}) & \alpha\hat{v}/\sqrt{\eta_{+}} & (\eta_{+}-\alpha)/\sqrt{\eta_{+}} & \hat{v}\\
(\eta_{-}-\alpha-\hat{v}^{2}) & \alpha\hat{v}/\sqrt{\eta_{-}} & (\eta_{-}-\alpha)/\sqrt{\eta_{-}} & \hat{v}\end{array}\right),\label{eq:Cinfty}\end{equation}
}For each value of $\alpha$ this problem is solved using a shooting
method. We start from the known solutions for $\hat{v}=0$ obtained
in \cite{Barashenkov89} and apply numerical continuation techniques
to locate the propagating pulses with $\hat{v}>0$. This procedure
for setting the boundary conditions at $\xi=L$, rather than simply
imposing $a(L)=0$, allows the shooting method to converge faster,
and the results obtained are found to be essentially independent of
$L$ for $L\gtrsim10$.

\begin{figure}[h]

\begin{raggedright}
\includegraphics[bb=0bp 0bp 703bp 324bp,scale=0.55]{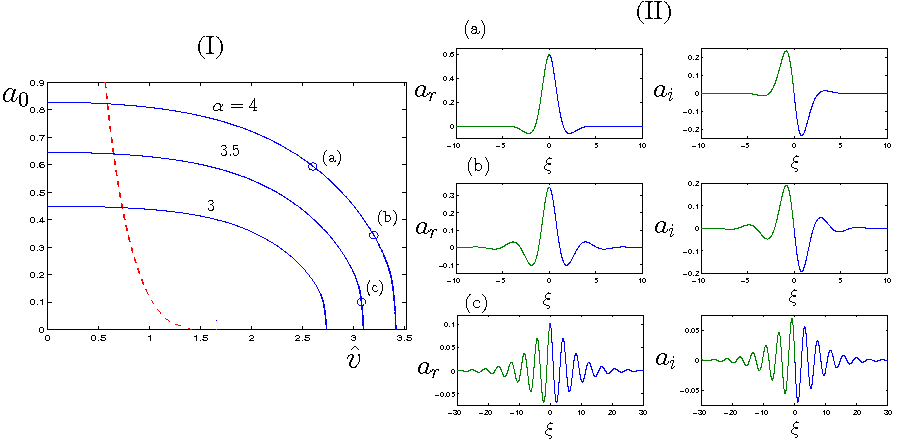}
\par\end{raggedright}

\caption{\label{fig:ar_homocla}(I) The solid lines correspond to homoclinic
cycles to the origin with $a_{0}>0$ for the indicated values of $\alpha$.
The dashed line bounds the regions where the origin is a saddle node
(on the left) and saddle-focus (on the right). (II) Spatial profiles
of three pulses for the values marked in (I) with open circles.}
\end{figure}

The left panel of Fig.~\ref{fig:ar_homocla} shows several families
of homoclinic orbits represented in the $(a_{0},\hat{v})$ plane for
different values of $\alpha$ and corresponding to the case $a_{0}>0$.
The dashed line separates the regions where the homoclinic orbit connects
to a saddle point, and where it connects to a saddle-focus, while
the open circles correspond to the solutions shown on the right panels.
In Fig.~\ref{fig:ar_homocla} II a-c the solutions can be seen to
develop oscillations as we move towards $a_{0}=0$. This corresponds
to moving along a horizontal line in Fig.~\ref{fig:eigval} (constant
$\alpha$) and to the right, approaching region (c) (precisely at
$a_{0}=0$, $\hat{v}=\sqrt{\alpha}+\sqrt{\alpha-2}$) where the eigenvalues
become purely imaginary. 

\begin{figure}[h]

\begin{raggedright}
\includegraphics[bb=0bp 0bp 703bp 324bp,scale=0.55]{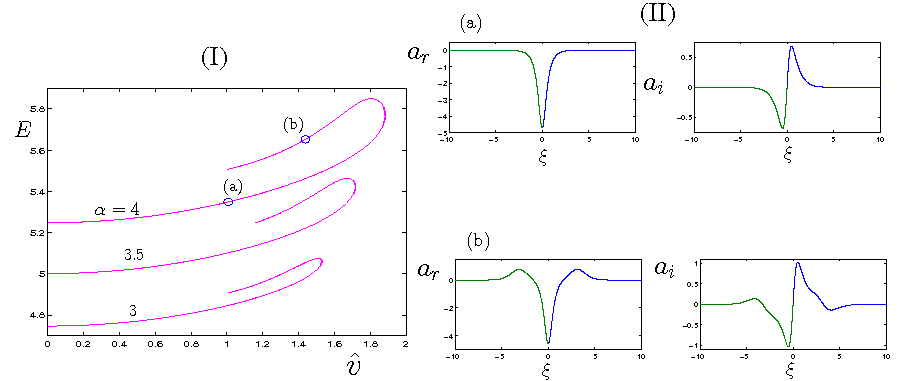}
\par\end{raggedright}

\caption{\label{fig:ar_homoclb} (I) Curves of homoclinic cycles to the origin
with $a_{0}<0$ for the indicated values of $\alpha$. (II) Spatial
profiles of two pulses corresponding to the values marked in (I) with
open circles.}
\end{figure}

This oscillatory behaviour, however, is not observed for the families
of pulse solutions found for $a_{0}<0$, as seen in Fig.~\ref{fig:ar_homoclb}.
Instead, these curves display turning points and, to better appreciate
these limit points, the results have been plotted in the plane $(E,\hat{v})$,
where $E$ is the positive quantity\[
E=\sqrt{\int_{0}^{\infty}\left(|a|^{2}+|a_{\xi}|^{2}\right)d\xi}.\]
In this case, as we move along the curves for fixed $\alpha$ and
past the turning point, the pulses develop two extra humps that tend
to move away from the origin (see Fig.~\ref{fig:ar_homoclb} II (a)
and (b)).

\section{NLCMEd simulations}

After having found a two parameter $(\alpha,\hat{v})$ family of symmetric
dispersive pulses (DP) that can be numerically continuated from the
solutions for $\hat{v}=0$ obtained in \cite{Barashenkov89}, we now
proceed to study their stability. 

\begin{figure}[t]
\begin{centering}
\includegraphics[bb=0bp 0bp 503bp 524bp,scale=0.73]{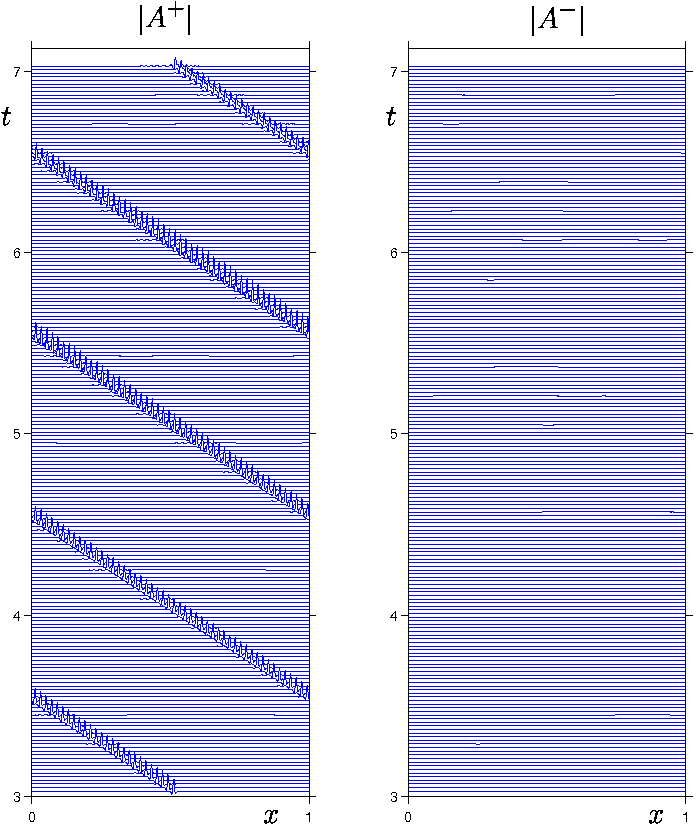}
\par\end{centering}

\caption{\label{fig:xtpinestv05}Space-time representation of the solution
of the NLCMEd ($\sigma=1/2$, $\kappa=1$, $\varepsilon=10^{-5}$,
and periodicity boundary conditions) for a unstable pulse propagating
on top of a CW ($\rho=1$ and $\theta=\pi/4$). The pulse parameters
are $\alpha=4$ and $\hat{v}=1$, and it corresponds to the point
labeled (a) in Fig.~\ref{fig:ar_homoclb}. See the file \texttt{movie3.1.gif}
for an animation of this pulse destabilization.}
\end{figure}

The idea is not to perform a complete stability analysis of the family
of DP but to show that stable DP can be found, and that the DP can
thus be considered as robust realizable localized structures of light
propagation in FBG. To do this we select several DP, place them on
top of their corresponding background CW, add a small random perturbation,
and use them as initial conditions for the full system of NLCMEd (\ref{eq:NLCMEd+}-\ref{eq:NLCMEd-})
that we numerically integrate for a certain amount of time with periodic
boundary conditions. The numerical method for the integration of the
NLCMEd uses a Fourier series in space with $N_{F}$ modes and a fourth
order Runge-Kutta scheme for the time integration of the resulting
system of ordinary differential equations for the Fourier coefficients.
The stiff linear diagonal terms associated with the small dispersion
coefficients are integrated implicitly, and the nonlinear terms are
computed in physical space with the usual 2/3 rule to avoid the aliasing
effects \cite{CanutoHussaniQuarteroniZang} (the maximum required
resolution for the simulations in this paper were $N_{F}=4096$ and
$\Delta t=.0005$). 

Unstable DP simply do not persist and their shape changes, as it can
be appreciated from the spatio-temporal evolution shown in Fig.~\ref{fig:xtpinestv05},
which corresponds to the pulse labeled (a) in Fig.~\ref{fig:ar_homoclb}
(see also the corresponding animation \texttt{movie3.1.gif}). Note
that the size of the pulse grows, at $t=5.5$ it is larger than at
$t=3$, and then it decays again at $t=7.0$. A few time units later
the oscillatory tails spread and the pulse structure is eventually
lost (not shown in the Figure). All DP explored for $a_{0}<0$ propagated
over the same simple CW with parameters $\rho=1$ and $\theta=\pi/4$
(cf. eqs. (\ref{CW+})-(\ref{CW-})), which corresponds to $\alpha=4$
in eq. (\ref{eq:eqsym}), and all were found to be unstable regardless
of its propagation speed $\hat{v}$. 

\begin{figure}[H]
\begin{centering}
\includegraphics[bb=0bp 0bp 503bp 324bp,scale=0.73]{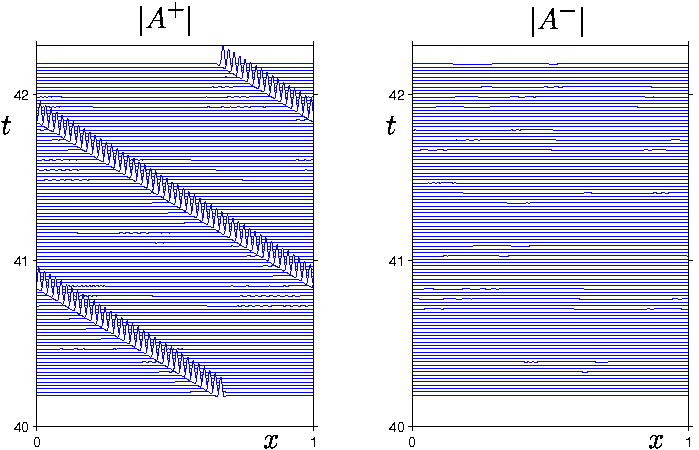}
\par\end{centering}

\caption{\label{fig:xtpestv13}Space-time representation of the solution of
the NLCMEd ($\sigma=1/2$, $\kappa=1$, $\varepsilon=10^{-5}$, and
periodicity boundary conditions) for a stable pulse propagating on
top of a CW ($\rho=1$ and $\theta=\pi/4$). The pulse parameters
are $\alpha=4$ and $\hat{v}=2.6$, and it corresponds to the point
labeled (a) in Fig.~\ref{fig:ar_homocla}. See the file \texttt{movie3.2.gif}
for an animation of this propagating pulse.}
\end{figure}

\begin{figure}[h]
\begin{centering}
\includegraphics[bb=0bp 0bp 503bp 324bp,scale=0.73]{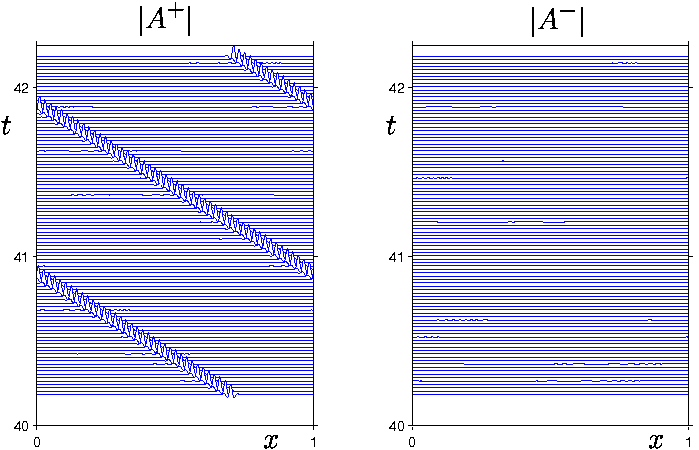}
\par\end{centering}

\caption{\label{fig:xtpestv16}Space-time representation of the solution of
the NLCMEd ($\sigma=1/2$, $\kappa=1$, $\varepsilon=10^{-5}$, and
periodicity boundary conditions) for a stable pulse propagating on
top of a CW ($\rho=1$ and $\theta=\pi/4$). The pulse parameters
are $\alpha=4$ and $\hat{v}=3.2$, and it corresponds to the point
labeled (b) in Fig.~\ref{fig:ar_homocla}. See the file \texttt{movie3.3.gif}
for an animation of this propagating pulse.}
\end{figure}

On the other hand, for $a_{0}>0$ and for the same background CW (i.e.,
$\alpha=4$), the DP are found to be unstable approximately for $0\le\hat{v}\lesssim2.2$
and stable for $v\gtrsim2.2$. The evolution of two stable pulses
is shown in Figs.~\ref{fig:xtpestv13} and \ref{fig:xtpestv16},
which correspond to $\hat{v}=2.6$ and $\hat{v}=3.2$, respectively,
where the structure of the slightly perturbed pulses is seen to remain
now virtually unaltered after more than $40$ time units (see also
the corresponding animations \texttt{movie3.2.gif} and \texttt{movie3.3.gif}). 

\begin{figure}
\begin{centering}
\includegraphics[bb=0bp 0bp 503bp 424bp,scale=0.73]{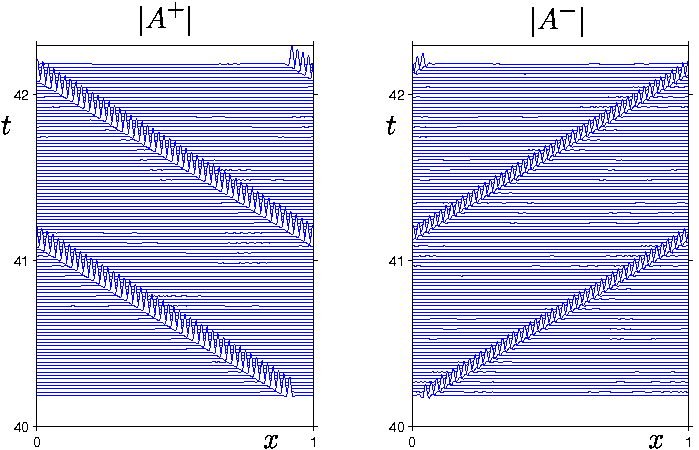}
\par\end{centering}

\caption{\label{fig:xtdospe}Space-time representation of the solution of
the NLCMEd ($\sigma=1/2$, $\kappa=1$, $\varepsilon=10^{-5}$, and
periodicity boundary conditions) showing the simultaneous propagation
in opposite directions of the two stable pulses from Figs. \ref{fig:xtpestv13}
and \ref{fig:xtpestv16}. See the file \texttt{movie3.4.gif} for an
animation of this pulse interaction.}
\end{figure}

Another very interesting feature of the DP that is worth mentioning
is the fact that they are somehow transparent to each other: two DP
propagating in opposite directions just pass through each other without
distortion. This is illustrated in Fig.~\ref{fig:xtdospe} (see also
the animation \texttt{movie3.4.gif}) where two stable DP (corresponding
to those in Figs.~\ref{fig:xtpestv13} and \ref{fig:xtpestv16})
are sent towards each other and after $40$ time units (approximately
$80$ collisions) they still remain practically undistorted. The reason
of this behavior is the dominant character of the transport effect
induced by the group velocity. If we rewrite the NLCMEd for a DP with
short (dispersive) spatial and temporal scales $X=x/\sqrt{\varepsilon}\sim1$
and $T=t/\sqrt{\varepsilon}\sim1$, they take the form of two uncoupled
wave equations\begin{eqnarray*}
A_{T}^{+}=A_{X}^{+}+\dots,\\
A_{T}^{-}=-A_{X}^{-}+\dots,\end{eqnarray*}
and then it is clear that DP travelling in opposite directions simply
propagate through different channels, and are (in first approximation)
completely independent. 

\section{Conclusions}

In this paper we have studied the effect of dispersion in the weakly
nonlinear dynamics of light propagation in a FBG. We have shown that
the (often negelected) small dispersion terms play a crucial role
in the transport dominated dynamics of light propagation in FBG. Dispersion
can give rise to complex spatio-temporal chaotic states, but also
to a new family of localized states (DP) that propagate over a CW
background and do not interact with each other. These DP are approximately
advected by the group velocity, but this transport effect does not
play any role in the determination of their internal structure, which
results basically from a balance between nonlinearity and dispersion.
It is also important to emphasize that this type of dynamics is not
contained in the standard dispersion-less NLCME formulation for light
propagation in FBG. Moreover, this behavior is just the result of
the competition of two effects with different asymptotic order: transport
and dispersion, and this is a generic situation that applies to any
propagative extended system unless some special care is taken to reduce
the group velocity (similar effects have been previously described
in the context of Hopf bifurcation in dissipative systems \cite{martelvega98}
and in parametrically forced surface waves \cite{martelvegaknobloch03}). 

\bibliographystyle{siam} 

\begin{thebibliography}{10}

\bibitem{Aceves00}
{\sc A.B. Aceves}, {\em Optical gap solitons: Past, present and future; theory
  and experiment}, CHAOS, 10 (2000), pp.~584--589.

\bibitem{Barashenkov96}
{\sc I.V. Barashenkov and Yu.~S. Smirnov}, {\em Existence and stability chart
  for the ac-driven, damped nonlinear schrodinger solitons}, Phys. Rev. E, 54
  (1996), pp.~5707--5725.

\bibitem{Barashenkov89}
{\sc I.V. Barashenkov, T.~Zhanlav, and M.M. Bogdan}, {\em Instabilities and
  soliton strucutres in the driven nonlinear schroedinger equation}, in
  Nonlinear World vol. 1, Proceedings of the IV International Workshop on
  Nonlinear and Turbulent Processes in Physics, V.G. Bar'yakhtar, V.M.
  Chernousenko, N.S. Erokhin, A.G. Sitenko, and V.E. Zakharov, eds., World
  Scientific, 1989.

\bibitem{CanutoHussaniQuarteroniZang}
{\sc C.~Canuto, H.Y. Hussani, A.~Quarteroni, and T.A. Zang}, {\em Spectral
  Methods in Fluid Mechanics}, Springer Series in Computational Physics,
  Springer-Verlag, 1988.

\bibitem{deSterke98}
{\sc C.M. de~Sterke}, {\em Theory of modulational instability in fiber bragg
  gratings}, J. Opt. Soc. Am. B, 15 (1998), pp.~2660--2667.

\bibitem{deSterkeSipe94}
{\sc C.M. de~Sterke and J.E. Sipe}, {\em Gap solitons}, Progress in Optics,
  XXXIII (1994), pp.~203--260.

\bibitem{GoodmanWeinsteinHolmes01}
{\sc R.H. Goodman, M.I. Weinstein, and P.J. Holmes}, {\em Nonlinear propagation
  of light in one-dimensional periodic structures}, J. Nonlinear Sci., 11
  (2001), pp.~123--168.

\bibitem{Kashyap99}
{\sc Raman Kashyap}, {\em Fiber Bragg Gratings}, Optics and Photonics, Academic
  Press, 1999.

\bibitem{Keller76}
{\sc H.B. Keller}, {\em Numerical solution of two point boundary value
  problems}, vol.~24 of CBMS-NSF Regional Conferences in Applied Mathematics,
  SIAM, 1976.

\bibitem{KevorkianCole96}
{\sc J.~Kevorkian and J.D. Cole}, {\em Multiple Scale and Singular Perturbation
  Methods}, vol.~114 of Applied Mathematical Sciences, Springer-Verlag, 1996.

\bibitem{Martel05}
{\sc C.~Martel}, {\em Dispersive destabilization of nonlinear light propagation
  in fiber bragg gratings}, CHAOS, 15 (2005), p.~013701.

\bibitem{MartelCasas06}
{\sc C.~Martel and C.M. Casas}, {\em Dispersive destabilization of nonlinear
  light propagation in fiber bragg gratings: a numerical verification}, CHAOS,
  17 (2007), p.~013114.

\bibitem{martelvega98}
{\sc C.~Martel and J.M. Vega}, {\em Dynamics of a hyperbolic system that
  applies at the onset of the oscillatory instability}, Nonlinearity, 11
  (1998), pp.~105--142.

\bibitem{martelvegaknobloch03}
{\sc C.~Martel, J.M. Vega, and E.~Knoboch}, {\em Dynamics of counterpropagating
  waves in parametrically driven systems: dispersion vs. advection}, Physica D,
  174 (2003), pp.~198--217.

\bibitem{SakaguchiMalomed04}
{\sc H.~Sakaguchi and B.A. Malomed}, {\em Dynamics of positive- and
  negative-mass solitons in optical lattices and inverted traps}, J. Phys. B:
  At. Mol. Opt. Phys., 37 (2004), pp.~1443--1459.

\bibitem{Wiggins88}
{\sc S.~Wiggins}, {\em Global Bifurcation and Chaos}, vol.~73 of Applied
  Mathematics Series, Springer-Verlag, 1988.

\bibitem{WinfulCooperman82}
{\sc H.G. Winful and G.D. Cooperman}, {\em Self-pulsing and chaos in
  distributed feedback bistable optical devices}, Appl. Phys. Lett., 40 (1982),
  pp.~298--300.

\bibitem{yulinskryabin03}
{\sc A.V. Yulin and D.V. Skryabin}, {\em Out-of-gap bose-einstein solitons in
  optical lattices}, Phys. Rev. E, 67 (2003), p.~023611.

\end{thebibliography}

\end{document}